\documentclass[
prc,%
10pt,%
final,%
notitlepage,%
oneside,%
twocolumn,%
nobibnotes,%
nofootinbib,
superscriptaddress,%
floatfix,%
showkeys,%
showpacs]%
{revtex4}

\usepackage{color} 
\usepackage{amsfonts} 
\usepackage{amsbsy} 
\usepackage{mathrsfs}
\usepackage{graphicx}
\usepackage{comment}

\def\lsim{\mathrel{\rlap{ \lower4pt\hbox{\hskip-3pt$\sim$}}
    \raise1pt\hbox{$<$}}} 
\def\gsim{\mathrel{\rlap{ \lower4pt\hbox{\hskip-3pt$\sim$}}
    \raise1pt\hbox{$>$}}} 
\def\scr#1{\mbox{\scriptsize #1}}

\begin{document}

\title{Light Nuclei Production in Au+Au Collisions at $\sqrt{s_{NN}}=$ 3 GeV
 within Thermodynamical Approach: Bulk Properties and Collective Flow}

\author{M. Kozhevnikova}\thanks{e-mail: kozhevnikova@jinr.ru}
\affiliation{Veksler and Baldin Laboratory of High Energy Physics,
  JINR Dubna, 141980 Dubna, Russia}
\author{Yu. B. Ivanov}\thanks{e-mail: yivanov@theor.jinr.ru}
\affiliation{Bogoliubov Laboratory of Theoretical Physics, JINR Dubna,
  141980 Dubna, Russia} 
  \affiliation{National Research Center
  "Kurchatov Institute", 123182 Moscow, Russia}

\begin{abstract}
We present results of simulations of light-nuclei production 
in Au+Au collisions at collision energy of $\sqrt{s_{NN}}=$ 3 GeV within updated
Three-fluid Hydrodynamics-based Event Simulator Extended 
by UrQMD (Ultra-relativistic Quantum Molecular Dynamics) final State interactions (THESEUS). 
The results are compared with recent STAR data. 
The light-nuclei production is treated within the thermodynamical approach on equal basis with hadrons.
The only additional parameter related to the light nuclei is the energy density of 
late freeze-out that imitates afterburner stage of the collision because the 
light nuclei do not participate in the UrQMD evolution. 
It is found that the late freeze-out is preferable for deuterons, tritons, and $^3$He.  
Remarkably, the $^4$He observables are better reproduced with the standard freeze-out. 
This suggests that the $^4$He nuclei better survive 
in the afterburner stage because they are more spatially compact and tightly bound objects. 
This is an argument in favor of dynamical treatment of light nuclei. 
The simulations indicate that the collision dynamics is determined by the hadronic phase. 
The calculated results reveal not perfect but a good reproduction of the data 
on  bulk observables and directed flow. 
The elliptic flow turns out to be more intricate.  
\end{abstract}
\maketitle

\section{Introduction}

During recent years the light-nuclei production has
become again one of the central topics in studies of relativistic heavy-ion collisions. 
As predicted, an enhanced production of light nuclei is a promising signal of the critical
endpoint (CEP) \cite{Shuryak:2019ikv,Shuryak:2020yrs,Sun:2020zxy}. 
This prediction is based on enhancement of the nucleon 
attraction near the CEP due to a rapid increase of correlation length \cite{Stephanov:1998dy}
and slowing down the equilibration of the density fluctuations \cite{Berdnikov:1999ph}
in the critical region. 
Abundant production of light nuclei also results from formation of baryon clusters due
to spinodal decomposition associated with mechanically unstable region in the first-order 
phase transition 
\cite{Skokov:2008zp,Skokov:2009yu,Randrup:2009gp,Steinheimer:2012gc,Steinheimer:2019iso,Sun:2022cxp}. 
These expectations revived interest to study of the light-nuclei production at high collision energies.  
At lower energies, a noticeable part of the baryon charge is emitted in the form of light nuclei. 
Therefore, even the proton data cannot be described without proper reproduction of the light-nuclei yield. 
There are different approaches to the light-nuclei production, which are still 
actively debated \cite{Mrowczynski:2020ugu,Motornenko:2021nds,Kittiratpattana:2022ety,Kireyeu:2023bye}.

Coalescence is the most popular approach, see, e.g.,  
\cite{Aichelin:1991xy,Russkikh:1993ct,Ivanov:2005yw,Liu:2019nii,Zhu:2015voa,Steinheimer:2012tb,Dong:2018cye,Sombun:2018yqh,Zhao:2020irc,Hillmann:2021zgj,Zhao:2021dka}, 
which however needs additional parameters for description of the light-nuclei yield. 
The recently developed transport models 
\cite{Oliinychenko:2018ugs,Staudenmaier:2021lrg,Aichelin:2019tnk,Glassel:2021rod,Bratkovskaya:2022vqi,Sun:2021dlz,Sun:2022xjr,Wang:2023gta} 
treat light nuclei microscopically on equal basis with other hadrons. 
However, these transport models also require an extensive additional input for the light-nuclei description.

The thermodynamical approach does not need any additional parameters for the light-nuclei treatment.  
It describes the light nuclei on equal basis with hadrons,
i.e. in terms of temperatures and chemical potentials. 
This approach was first realized within the statistical model, 
which fairly well described
deuteron midrapidity yields \cite{STAR:2019sjh,STAR:2022hbp} at the energies from 7.7 to 200 GeV
\cite{Andronic:2010qu,Vovchenko:2020dmv}  
while overestimated the tritium yield 
by roughly a factor of two \cite{Vovchenko:2020dmv,Zhang:2020ewj}.
The statistical model gave a good description of  
even hypernuclei and antinuclei \cite{Andronic:2017pug}.

Inspired by relative success of the statistical model 
\cite{Andronic:2010qu,Vovchenko:2020dmv,Zhang:2020ewj,Andronic:2017pug}, 
we implemented the thermodynamic approach to the 
light-nuclei production into the updated THESEUS event generator \cite{Kozhevnikova:2020bdb}.
In general, this approach involves no additional parameters inherent in light nuclei, which makes 
its predictive power the same for light nuclei and hadrons. 
However, because of the lack of the (post-hydrodynamical) afterburner stage for the light nuclei 
we had to introduce a parameter of the late freeze-out for them \cite{Kozhevnikova:2022wms}. 
This late freeze-out imitates the afterburner evolution. 
In Ref. \cite{Kozhevnikova:2022wms} we considered Pb+Pb and Au+Au collisions in the collision energy range 
of $\sqrt{s_{NN}}=$  6.4--19.6 GeV.  
The updated THESEUS resulted in an imperfect but reasonable reproduction of data on bulk observables
of the light nuclei, especially their functional dependence on the collision energy and light-nucleus mass.

Data on light-nuclei production in Au+Au collisions at $\sqrt{s_{NN}}=$ 3 GeV were recently
published \cite{STAR:2023uxk,STAR:2021ozh}. 
Apart from several Blast-Wave fits, these data on bulk properties of light nuclei 
were analyzed in coalescence-based 3D simulations within 
JAM (Jet AA microscopic transport Model \cite{Nara:1999dz,Isse:2005nk}), SMASH  
(Simulating Many Accelerated Strongly-interacting Hadrons \cite{SMASH:2016zqf}), 
and UrQMD \cite{Bass:1998ca}, which were presented in the
STAR paper \cite{STAR:2023uxk}, as well as in the JAM-based calculation \cite{Xu:2023xul}.  
Simulations were also performed within the PHQMD (Parton-Hadron-Quantum-Molecular-Dynamics) 
approach \cite{Bratkovskaya:2022vqi}, 
in which the cluster formation occurs dynamically due to interactions, and 
the hybrid dynamical-statistical approach \cite{Buyukcizmeci:2023azb}. 
These models reproduce, albeit to varying degrees, experimental trends of the data.  
The data on the collective flow of light nuclei at 3 GeV \cite{STAR:2021ozh} were analyzed within 
the coalescence-based JAM generator, presented in Ref. \cite{STAR:2021ozh}. 
These JAM simulations well describe the data.

In the present paper, the treatment of our previous paper \cite{Kozhevnikova:2022wms} is extended to Au+Au collisions at 
$\sqrt{s_{NN}}=$ 3 GeV. We calculate bulk properties, directed and elliptic flows of protons and light nuclei 
($d$, $t$, $^3$He and $^4$He) within the updated THESEUS approach \cite{Kozhevnikova:2020bdb} 
based on thermodynamic treatment of light nuclei. In contrast to the collision energy range considered 
in Ref. \cite{Kozhevnikova:2022wms}, the yield of the light nuclei at 3 GeV plays noticeable role in 
the total balance of the baryon charge.

\section{updated THESEUS} 
  \label{updated THESEUS}

The THESEUS event generator \cite{Batyuk:2016qmb,Batyuk:2017sku}
is based on the  model of the three-fluid dynamics (3FD) \cite{Ivanov:2005yw,Ivanov:2013wha}
complemented by the UrQMD~\cite{Bass:1998ca} for the afterburner stage.
The 3FD  takes into account counterstreaming of the leading baryon-rich matter at the early stage of
nuclear collisions. This nonequilibrium stage is modeled 
by means of two counterstreaming baryon-rich fluids.  
Newly produced particles, which dominantly populate the midrapidity region,
are assigned to a so-called fireball fluid.  These fluids are governed  
by set of hydrodynamic equations coupled by friction
terms, which describe the energy--momentum exchange between the fluids.

The output of the 3FD model, i.e. the freeze-out hypersurface, is recorded in terms of
local flow velocities and thermodynamic quantities. 
The THESEUS generator transforms the 3FD output
into a set of observed particles, i.e. performs the particlization. 
After the particlization the afterburner stage is described by the UrQMD. 
First applications of the THESEUS generator to description of 
heavy-ion collisions were demonstrated in Refs.~\cite{Batyuk:2016qmb,Batyuk:2017sku}.

In the initial version of the THESEUS \cite{Batyuk:2016qmb,Batyuk:2017sku},  
spectra of the so-called primordial nucleons, 
i.e. both observable nucleons and those bound in the light nuclei, were calculated. 
These spectra were intended for 
the subsequent use in the coalescence model  \cite{Ivanov:2005yw,Ivanov:2017nae} 
for the light-nuclei production. 
Therefore, the nucleons bound in the light nuclei should be  
subtracted from the primordial ones in order to obtain the observable nucleons. 
Such subtraction is performed in the 3FD, where production of the light nuclei is calculated within the coalescence approach \cite{Ivanov:2005yw,Ivanov:2017nae}. The initial version of the 
THESEUS took temperature and chemical potential fields for hadron sampling from the hydrodynamic output 
of the 3FD (where the clusters are not included in the EoS), and produced both hadrons and clusters within the thermodynamical approach. 
This led to an overestimation of the total baryon charge in the final state
containing both baryons and clusters. 
Therefore, a compensating correction was required. 
Such correction was made in the updated version of THESEUS \cite{Kozhevnikova:2020bdb}
by means of the recalculation of the baryon chemical potentials, proceeding from the local
baryon number conservation in the system of hadrons extended by
the light-nuclei species listed in Tab. \ref{tab:clusters}.
The list of the light nuclei includes the stable nuclei  
[deuterons ($d$), tritons ($t$), helium isotopes $^3$He and $^4$He], 
and low-lying $^4$He resonances decaying into  stable species  \cite{Shuryak:2019ikv}. 
The corresponding anti-nuclei are also included. 
\begin{table}[ht]
\begin{center}
\begin{tabular}{|c|c|c|}
\hline
Nucleus($E$[MeV]) & $J$ &  decay modes, in \% \\
\hline
\hline
$d$           & $1$ & Stable \\
$t$           & $1/2$ & Stable \\
$^3$He        & $1/2$ & Stable \\
$^4$He        & $0$ & Stable \\
$^4$He(20.21) & $0$ & $p$ = 100\\
$^4$He(21.01) & $0$ & $n$ = 24,  $p$ = 76\\
$^4$He(21.84) & $2$ & $n$ = 37,  $p$ = 63  \\
$^4$He(23.33) & $2$ & $n$ = 47,  $p$ = 53  \\
$^4$He(23.64) & $1$ & $n$ = 45,  $p$ = 55 \\
$^4$He(24.25) & $1$ & $n$ = 47,  $p$ = 50,  $d$ = 3 \\
$^4$He(25.28) & $0$ & $n$ = 48,  $p$ = 52\\
$^4$He(25.95) & $1$ & $n$ = 48,  $p$ = 52 \\
$^4$He(27.42) & $2$ & $n$ = 3,   $p$ = 3,   $d$ = 94 \\
$^4$He(28.31) & $1$ & $n$ = 47,  $p$ = 48,  $d$ = 5 \\
$^4$He(28.37) & $1$ & $n$ = 2,   $p$ = 2,   $d$ = 96 \\
$^4$He(28.39) & $2$ & $n$ = 0.2, $p$ = 0.2, $d$ = 99.6 \\
$^4$He(28.64) & $0$ & $d$ = 100 \\
$^4$He(28.67) & $2$ & $d$ = 100 \\
$^4$He(29.89) & $2$ & $n$ = 0.4, $p$ = 0.4, $d$ = 99.2 \\
\hline
\end{tabular}
\caption{Stable light nuclei and low-lying resonances of the~$^4$He system (from BNL properties of nuclides 
\cite{www.nndc.bnl}).
$J$ denotes the total angular momentum. 
The last column represents branching ratios of the decay channels, in per cents. The 
$p,n,d$ correspond to the emission of protons, neutrons, or deuterons, respectively.
}
\label{tab:clusters}
\end{center}
\end{table}
\begin{figure*}[!tbh]
  \includegraphics[width=.7\textwidth]{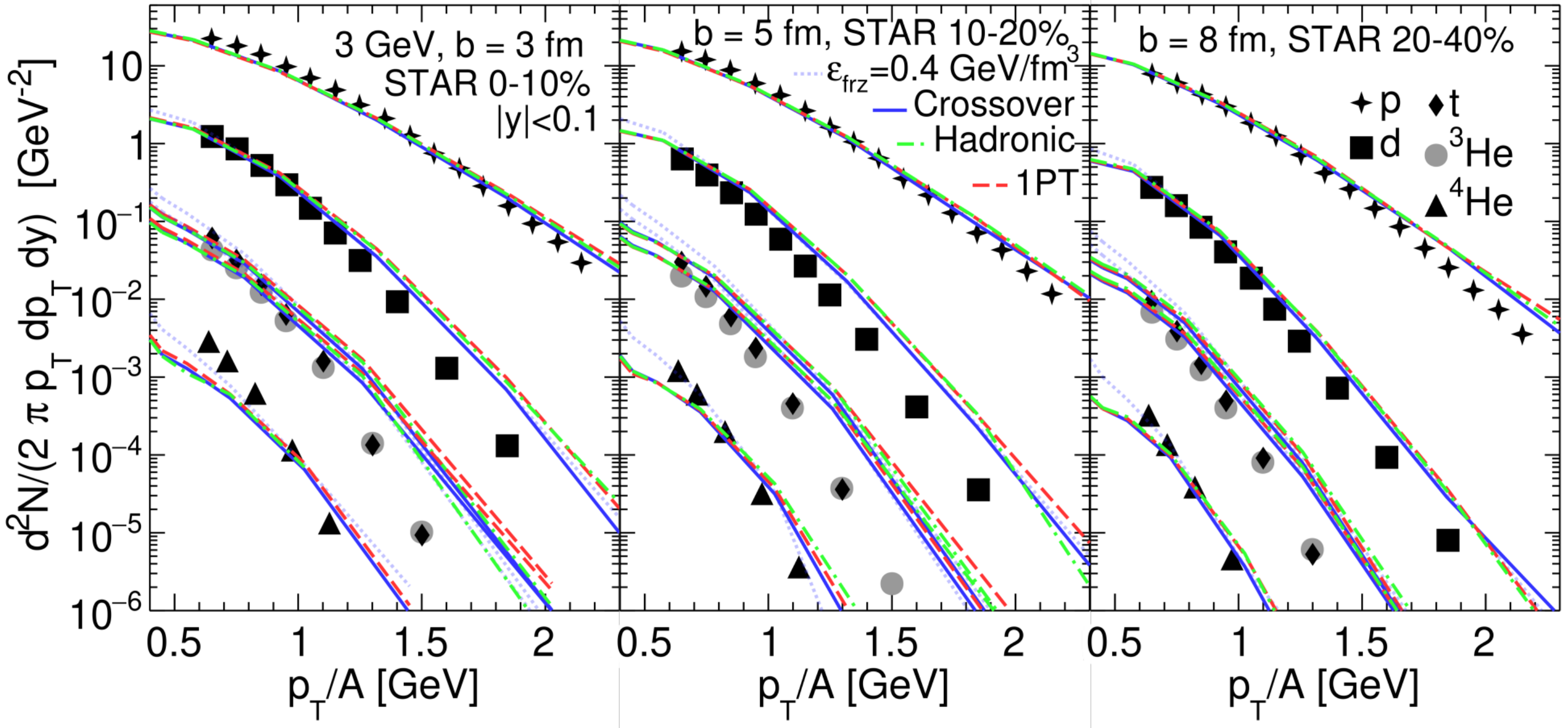}
  \caption{
Midrapidity ($|y|<$ 0.1) transverse-momentum spectra of protons and light nuclei 
(deuterons, tritons, $^3$He and $^4$He)  in  
Au+Au collisions at collision energy of $\sqrt{s_{NN}}=$ 3 GeV  and
different centralities (impact parameters $b$). Results are calculated 
with hadronic, 1PT and crossover EoS's. 
Results of the THESEUS simulations with the 
late freeze-out ($\varepsilon_{\rm frz}=$ 0.2 GeV/fm$^3$) for three EoS's and the  
conventional 3FD freeze-out ($\varepsilon_{\rm frz}=$ 0.4 GeV/fm$^3$) for the crossover EoS 
are displayed for light nuclei. 
Protons are calculated within the conventional 3FD freeze-out with the subsequent UrQMD afterburner.
STAR data are from Ref. \cite{STAR:2023uxk}.
}	
    \label{fig:ptA_3GeV_frz02_frz04-all}
\end{figure*}

In the updated version of the THESEUS 
\cite{Kozhevnikova:2020bdb}, the light nuclei were included 
on equal basis with other hadrons. 
These nuclei are sampled similarly to other hadrons, i.e. accordingly to their 
phase-space distribution functions.
However, there is an important difference. 
While the hadrons
pass through the UrQMD afterburner stage after the particlization, the light nuclei do not,  
because the UrQMD is not able to treat them. This is a definite shortcoming because the light nuclei
are destroyed and again re-produced during the afterburner stage 
\cite{Oliinychenko:2018ugs,Staudenmaier:2021lrg,Glassel:2021rod,Sun:2021dlz,Sun:2022xjr}. 
Following the recipe of Ref. \cite{Kozhevnikova:2020bdb}, we imitate the afterburner for light nuclei by 
late freeze-out in the 3FD.

Three different equations of state (EoS's) are used
in the 3FD simulations: 
a purely hadronic EoS \cite{gasEOS} (hadr. EoS) and two EoS's
with deconfinement \cite{Toneev06}, i.e. 
an EoS with a first-order phase transition (1PT EoS) and one with a
smooth crossover transition (crossover EoS).  
Consequently, the 3FD output for these EoS's is used in the THESEUS generator.

\section{Bulk Observables} 
  \label{Bulk Observables}

To partially overcome the aforementioned problem of the afterburner stage for the light nuclei,
we imitate the afterburner effect by late freeze-out for light nuclei. Similarly to Ref. 
\cite{Kozhevnikova:2020bdb}, we take the freeze-out energy density $\varepsilon_{\rm frz}=$ 0.2 GeV/fm$^3$
for this late freeze-out, which looks quite suitable for all considered quantities, as seen below. 
No other additional tuning of the parameters was carried out for the light nuclei.
Note that the conventional 3FD freeze-out energy density for all other hadrons, subjected to 
the UrQMD afterburner, is 0.4 GeV/fm$^3$. Details of the freeze-out procedure in the 3FD are described in Refs.
\cite{Russkikh:2006aa,Ivanov:2008zi}.  
The $\varepsilon_{\scr{frz}}$ quantity has meaning of a ``trigger'' 
that indicates possibility of the freeze-out.
The freeze-out procedure begins when the local (i.e. in a cell) energy density drops below the freeze-out value 
$\varepsilon_{\scr{frz}}$, then testing for additional freeze-out conditions starts.
If all the freeze-out conditions are met, the cell is declared frozen-out.  
Thus, the actual energy density of a frozen-out cell turns out to be lower than $\varepsilon_{\scr{frz}}$,  
as it is demonstrated, e.g., in Ref. \cite{Ivanov:2020wak}.

\subsection{Transverse-momentum spectra} 
  \label{Transverse-momentum spectra}

\begin{figure*}[!tbh]
  \includegraphics[width=.93\textwidth]{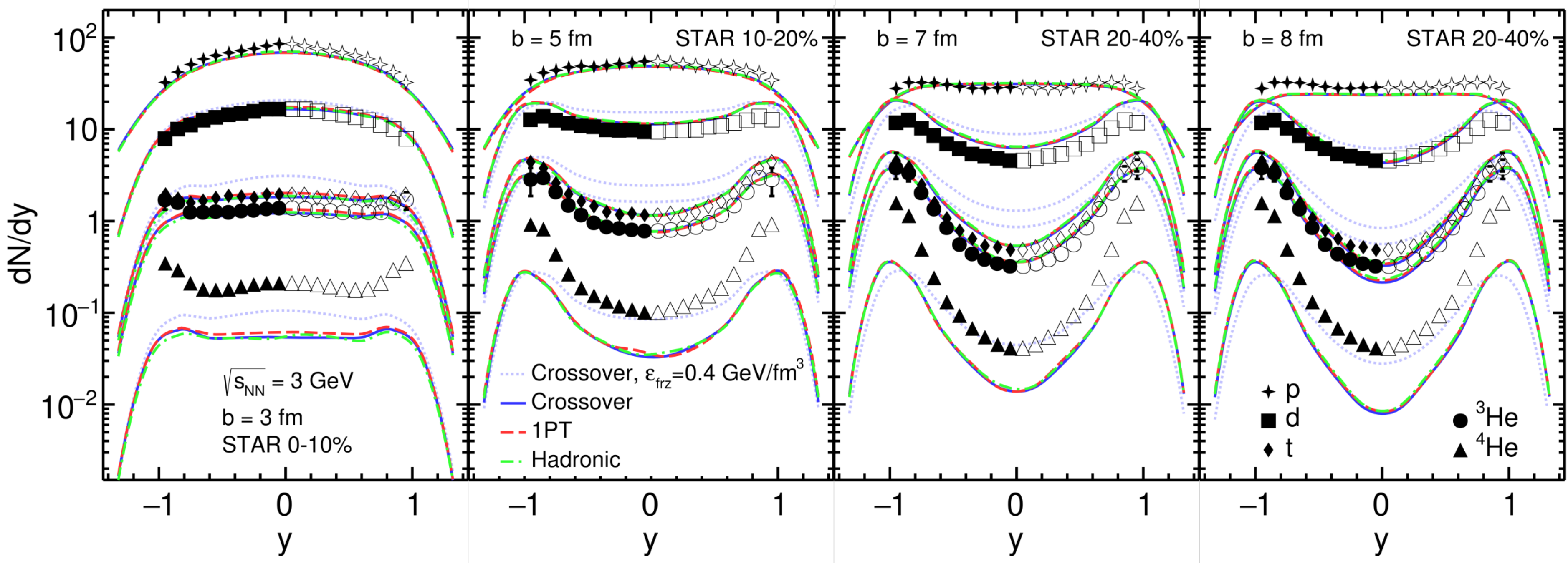}
  \caption{
Rapidity distributions of protons and light nuclei 
(deuterons, tritons, $^3$He and $^4$He)  in  
Au+Au collisions at collision energy of $\sqrt{s_{NN}}=$ 3 GeV  and
different centralities (impact parameters $b$). Results are calculated 
with hadronic, 1PT and crossover EoS's. 
Results of the THESEUS simulations with the 
late freeze-out ($\varepsilon_{\rm frz}=$ 0.2 GeV/fm$^3$) for three EoS's and the 
conventional 3FD freeze-out ($\varepsilon_{\rm frz}=$ 0.4 GeV/fm$^3$) for the crossover EoS 
are displayed for light nuclei. 
Protons are calculated within the conventional 3FD freeze-out with the subsequent UrQMD afterburner.
STAR data are from Ref. \cite{STAR:2023uxk}.
Full symbols display measured experimental points, whereas the open ones are those reflected with respect
to the midrapidity.}	
    \label{fig:dNdy_All_3AGeVmix_tph_gas_frz02_frz04}
\end{figure*}

Midrapidity ($|y|<$ 0.1) transverse-momentum spectra of protons and light nuclei 
(deuterons, tritons, $^3$He and $^4$He)  in  
Au+Au collisions at collision energy of $\sqrt{s_{NN}}=$ 3 GeV  and
different centralities (impact parameters $b$) are displayed in Fig. \ref{fig:ptA_3GeV_frz02_frz04-all}. 
The proton spectra are calculated within full THESEUS, i.e. with the standard 3FD freeze-out and the UrQMD 
afterburner. The spectra of light nuclei are evaluated for the late 3FD freeze-out without the 
afterburner stage. 
As seen, the results for different EoS's are almost identical, which means that the dynamics 
is dominated by the hadronic phase. The difference between the late freeze-out and the  
conventional 3FD freeze-out for the crossover EoS is mostly seen at low $p_T$ for light nuclei. 
In the $p_T$ spectra, this difference does not look dramatic. However, in rapidity distributions, 
Fig.  \ref{fig:dNdy_All_3AGeVmix_tph_gas_frz02_frz04}, which mostly determined by the low-$p_T$
spectra, the difference is quite noticeable.

Similarly to that found previously \cite{Kozhevnikova:2022wms}, the experimental $p_T$ spectra
\cite{STAR:2023uxk} turn out to be steeper than the calculated ones. This is not 
only the problem of light-niclei description. The proton spectra are also more flat than the 
experimental ones. This is a shortcoming of the 3FD model. The slight extra flatness of the 
proton spectra transforms into larger extra flatness of light-nuclei spectra.

As has been already noted in Ref. \cite{Kozhevnikova:2022wms},
the 3FD predictions overestimate the high-$p_T$ ends of the spectra
because of finiteness of the considered system. 
Even abundant hadronic probes become rare at high momenta. 
For the rare probes,  treatment on the basis of the canonical ensemble is needed 
rather than within the grand canonical ensemble. 
The grand canonical ensemble results in overestimation of their yields. 
Of course, it is difficult to indicate how much of this overestimation is due to the grand canonical treatment, and not to the shortcomings of the model.

\subsection{Rapidity distributions} 
  \label{Rapidity distributions}

Rapidity distributions of protons and light nuclei 
in  Au+Au collisions at collision energy of $\sqrt{s_{NN}}=$ 3 GeV  and
different centralities are presented in Fig. \ref{fig:dNdy_All_3AGeVmix_tph_gas_frz02_frz04}. 
Again the proton distributions are calculated within full THESEUS, i.e. with the standard 3FD freeze-out and the UrQMD 
afterburner, while the light-nuclei distributions, within the late 3FD freeze-out and without the 
afterburner stage. 
The light-nuclei distributions, calculated at the conventional 3FD freeze-out, 
are also displayed for comparison.  
The same value of the late-freeze-out energy density ($\varepsilon_{\rm frz}=$ 0.2 GeV/fm$^3$)
as that  at higher collision energies \cite{Kozhevnikova:2022wms} turned out to be the most suitable at 3 GeV. 
The reproduction of the experimental distributions turns out to be even better 
than that at higher collision energies \cite{Kozhevnikova:2022wms}. 
The THESEUS simulations well describe difference in the form of proton and light-nuclei distributions and its 
dependence on the centrality.

For the experimental centrality of 20-40\% we present comparison with results for 
two impact parameters ($b=$ 7 and 8 fm) 
in order to illustrate the sensitivity of the results to the choice of $b$. As seen, the proton rapidity 
density is underestimated in midrapidity at $b=$ 8 fm in spite of perfect reproduction of the low-$p_T$ experimental spectrum, 
see  Fig. \ref{fig:ptA_3GeV_frz02_frz04-all}. The reason is that the extrapolation of the experimental spectrum 
to even lower $p_T$ exceeds the THESEUS predictions. Similar situation takes place for the light nuclei.  
Thus, the results for two impact parameters ($b=$ 7 and 8 fm) illustrate uncertainty of the THESEUS predictions.

The $^4$He distributions deserve a separate discussion. 
The late-freeze-out calculation strongly underestimates these distributions.  
Expanding the list of light-nuclei resonances by those of $^5$H, $^5$He, and $^5$Li \cite{Vovchenko:2020dmv}, 
which decay into $^4$He, makes an additional contribution to the $^4$He yield. 
This additional contribution is large, i.e. of the order of 60\%, 
in central collisions at the energy of 3 GeV, accordingly to Ref. \cite{Vovchenko:2020dmv}.  
However, it is not large enough to compensate the obtained underestimation. 
At the same time, the standard-freeze-out calculation results in much better (almost perfect in midrapidity 
at 10-20\% and 20--40\% centralities) reproduction of the data. 
The $p_T$ spectra are also much better described by the standard freeze-out, 
see Fig. \ref{fig:ptA_3GeV_frz02_frz04-all}. 
This suggests that the $^4$He nuclei better survive 
in the afterburner stage because they are more spatially compact and tightly bound objects. 
Then the standard freeze-out is more relevant for their description. 
Note that the feed-down contribution to the $^4$He yield, $\approx$60\% \cite{Vovchenko:2020dmv}, 
is quite enough to drastically improve the reproduction midrapidity data in central collisions
by the standard-freeze-out calculation. 

At lower collision energies, the enhancement of the $^4$He production is even more 
spectacular \cite{FOPI:2010xrt}. In central Au+Au collisions, the $^4$He and $^3$He yields are approximately equal 
at $E_{lab} =$ 0.4$A$ GeV and the $^4$He yield even exceeds that of $^3$He at 0.15$A$ GeV. 
It seemingly contradicts to the thermodynamic picture of the light-nuclei production. 
However, this contradiction is removed, if the chemical freeze-out occurs earlier for $^4$He nuclei than 
for $d$, $t$ and $^3$He because of larger binding energy of the $^4$He nuclei. 
In Ref. \cite{Wang:2023gta} it is formulated in terms of the Mott transition \cite{Typel:2009sy}: 
The observed enhancement of
the $^4$He yield can be attributed to the weaker Mott effect on $^4$He nuclei than that
on deuteron, triton and $^3$He, as a result of its much larger binding energy. The cutoff value for 
the average nucleon phase-space density, $f_A^{\rm cut}$, that is used in Ref. \cite{Wang:2023gta}, 
see Eq. (5) in Ref. \cite{Wang:2023gta}, in fact plays the role of the effective freeze-out for 
different light nuclei.  As found in Ref. \cite{Wang:2023gta}, the $f_{A=4}^{\rm cut}$ value is 
approximately twice as large as that for lighter nuclei. This is consistent with our conclusion 
about the earlier freeze-out of $^4$He. 
In particular, the results of the kinetic approach of Ref. \cite{Wang:2023gta} imply that the freeze-out 
parameters for each light-nuclei specie are individual and depend on the binding energy of considered 
nucleus.

\subsection {Medium Effects} 
  \label{Medium Effects}

Study of the in-medium effects in light-nuclei production has started long ago \cite{Danielewicz:1992mi}. 
Later, coupled quantum kinetic equations were derived which describe the time evolution of the Wigner
distribution functions for nucleons and light clusters \cite{Kuhrts:2000zs}.
An alternative approach has also been proposed 
within the antisymmetrzed molecular dynamics (AMD)
approach \cite{Ono:2016mqk}, in which nucleons are represented in terms
of quantal wave packets, which are antisymmeztrized with each other. 
These approaches have been successfully applied 
to analysis of results of the GANIL experiment at 50$A$ MeV.

\begin{figure}[!tbh]
  \includegraphics[width=.33\textwidth]{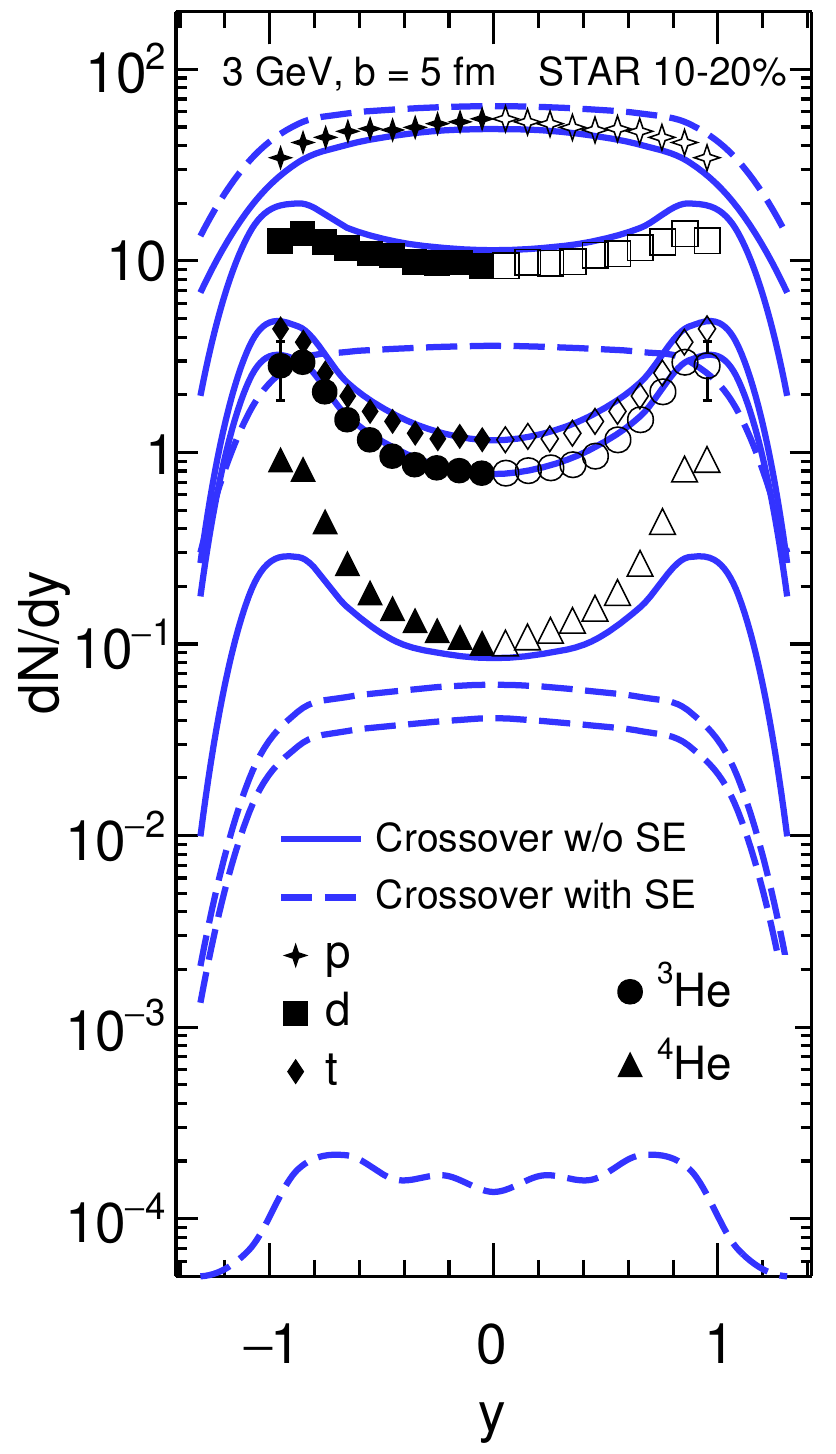}
  \caption{
Rapidity distributions of protons and light nuclei 
(top down: protons, deuterons, tritons, $^3$He and $^4$He)  in  
Au+Au collisions at collision energy of $\sqrt{s_{NN}}=$ 3 GeV  and $b=$ 5 fm.
Results are calculated in the crossover scenario with (with SE) 
and without (w/o SE) the self-energy contributions,  
i.e. the $\Delta E^{\rm SE}_A({\bf p})$ and $\Delta E^{\rm Pauli}_A({\bf p})$ 
terms in Eq. (\ref{SE}). 
Protons are calculated within the conventional 3FD freeze-out with the subsequent UrQMD afterburner.
Deuterons, tritons, and $^3$He are calculated with the 
late freeze-out ($\varepsilon_{\rm frz}=$ 0.2 GeV/fm$^3$), while $^4$He, the   
conventional 3FD freeze-out ($\varepsilon_{\rm frz}=$ 0.4 GeV/fm$^3$). 
STAR data are from Ref. \cite{STAR:2023uxk}.
Full symbols display measured experimental points, whereas the open ones are those reflected with respect
to the midrapidity.}	
    \label{fig:dNdy_All_3AGeVmix_5fm_oldSE}
\end{figure}

In Refs. \cite{Bastian:2016xna,Ropke:2017dur}, it was proposed to extend the study of these in-medium effects
to nuclear collisions at NICA (Nuclotron-based Ion Collider Facility) and FAIR (Facility for Antiproton and Ion Research) energies. A suitable frame for such a study is the thermodynamic approach to light-nuclei production
complemented by quantum statistical approach 
that includes medium effects 
due to Pauli blocking and self-energies \cite{Typel:2009sy,Ropke:2012qv,Ropke:2014fia}. 
This quantum statistical approach is based on the relativistic mean-field model of Ref. \cite{Ropke:2012qv}. 
The energies of light nuclei are given by the formula
\begin{eqnarray}
\label{SE}
E_A({\bf p}) = E^0_A({\bf p}) + \Delta E^{\rm SE}_A({\bf p}) + \Delta E^{\rm Pauli}_A({\bf p})
\end{eqnarray}
where $E^0_A(p)$ is the vacuum energy of $A$-nucleus with momentum ${\bf p}$, 
$\Delta E^{\rm SE}_A(p)$ is in-medium self-energy shift, 
$\Delta E^{\rm Pauli}_A(p)$ is the energy correction due to the Pauli blocking. 
The last two quantities also depend on the baryon density, temperature and proton/neutron asymmetry of
the matter. More details can be found in Ref. \cite{Ropke:2014fia}. 
This quantum statistical approach was incorporated into THESEUS at the particlization stage.

Preliminary results showed 
\cite{Blaschke:2020gqr,Blaschke:2020jmv}
that the best description of the light-nuclei yields is obtained
when the self-energy effects are discarded at few-GeV collision energies ($\sqrt{s_{NN}}$).  
However, those results were not conclusive because the proper conservation of the baryon charge 
was not achieved in that version of the THESEUS, see sect.  \ref{updated THESEUS}. 
The Pauli-blocking effects were disregarded in those calculations.

The present calculation are performed within the updated version of the THESEUS, 
where the conservation of the baryon charge is strictly fulfilled. 
An upper estimate of the Pauli-blocking effect is also made by means of its approximation by that at zero light-nucleus momentum in the rest frame the medium. 
The Pauli-blocking is strongest at this zero momentum. 
The Pauli-blocking effect turned out to be negligibly small because of  
high freeze-out temperatures as compared with  
the corresponding Fermi energies.  
The results of this calculation with the crossover EoS are displayed in 
Fig. \ref{fig:dNdy_All_3AGeVmix_5fm_oldSE}. 
As seen, the medium effect accordingly to Ref. \cite{Ropke:2014fia} turns out to be too strong. 
It results in strong disagreement with data, as has been already seen from the preliminary simulations  
\cite{Blaschke:2020gqr,Blaschke:2020jmv}. 
Even the proton yield is overestimated. 
Apparently this is because the 
relativistic mean-field model \cite{Ropke:2012qv}, underlying these medium corrections, 
has been parametrized to reproduce low-energy nuclear phenomena. This parametrization is simply inapplicable 
to highly excited nuclear matter. Of course, we could introduce a temperature-dependent attenuation factor 
to reduce the strength of these medium corrections. However, it would be a purely phenomenological tuning 
parameter. Therefore, we avoid doing this.

\section{Collective flow} 

Collective flow is a more subtle observable of the heavy-ion collisions. 
Its calculation with the thermodynamic approach is straightforward 
because light nuclei are treated on the equal basis with other hadrons. 
Results for the collective flow, which are presented below, are calculated with 
respect to the reaction plane that is exactly defined in the simulations. We have also done 
calculations with the event plane determined after the afterburner 
in terms of observable particles within the STAR acceptance. The reaction-plane and 
event-plane results turned out to be practically identical.

\subsection{Directed flow} 
  \label{Directed flow}

\begin{figure}[!tbh]
\includegraphics[width=.42\textwidth]{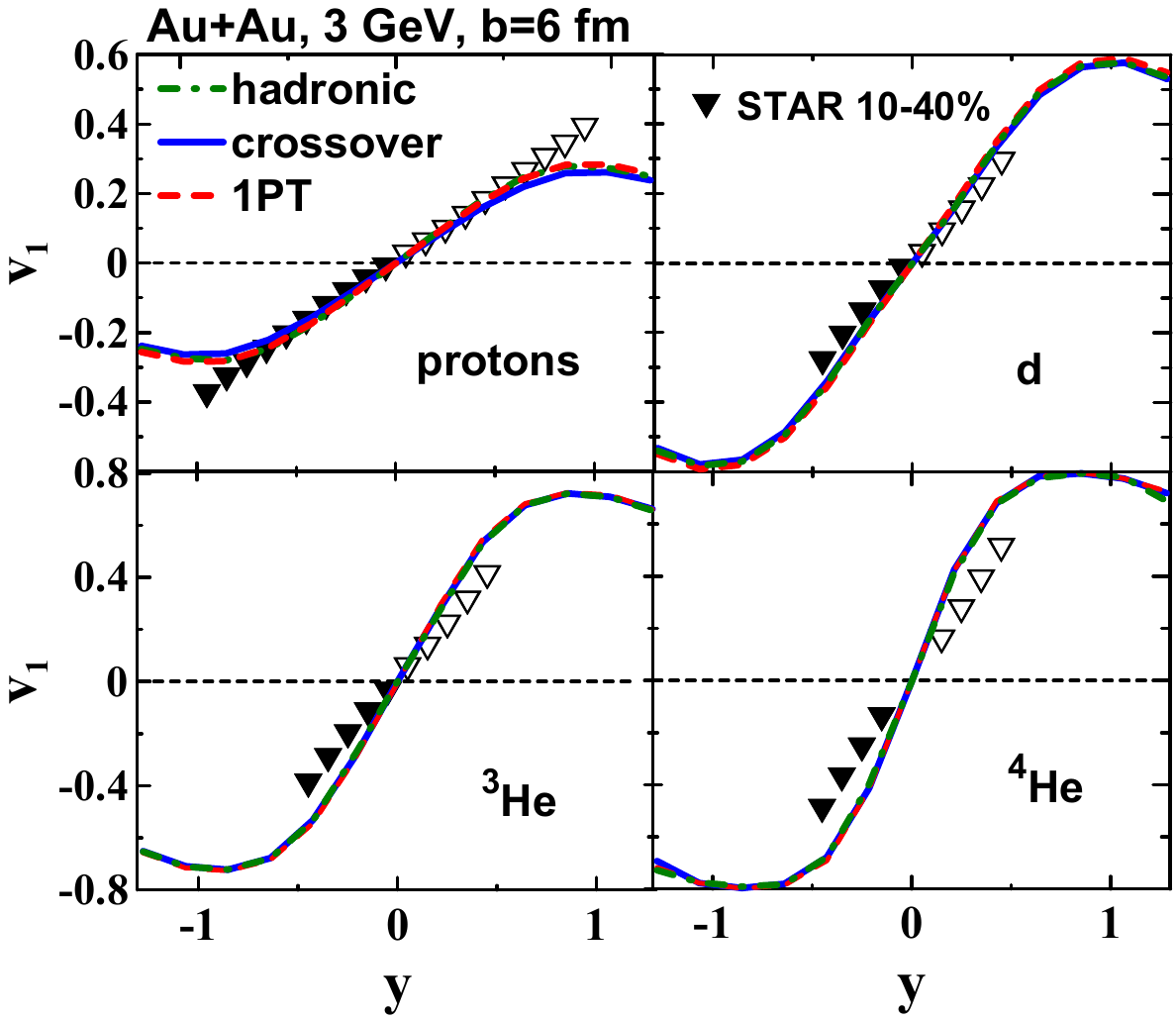}
  \caption{
Directed flow of  
protons and light nuclei 
(deuterons, $^3$He and $^4$He) as function of rapidity in semicentral ($b=$ 6 fm)    
Au+Au collisions at collision energy of $\sqrt{s_{NN}}=$ 3 GeV. 
Results are calculated 
with hadronic, 1PT and crossover EoS's. 
Results of the THESEUS simulations with the 
late freeze-out ($\varepsilon_{\rm frz}=$ 0.2 GeV/fm$^3$)  
are displayed for light nuclei. 
Protons are calculated within the conventional 3FD freeze-out with the subsequent UrQMD afterburner.
STAR data are from Ref. \cite{STAR:2021ozh,STAR:2021yiu}.
Full symbols display measured experimental points, whereas the open ones are those reflected with respect
to the midrapidity.}
\label{fig:v1-STAR-fxt-3GeV-fragments_frz02}
\end{figure}

The calculated directed flow of protons and light nuclei (deuterons, $^3$He and $^4$He) 
as function of rapidity in semicentral ($b=$ 6 fm)    
Au+Au collisions at collision energy of $\sqrt{s_{NN}}=$ 3 GeV is presented in 
Fig. \ref{fig:v1-STAR-fxt-3GeV-fragments_frz02}. 
The results are compared with STAR data \cite{STAR:2021ozh,STAR:2021yiu}.
We do not display results for the tritons because they are very similar to those for $^3$He, 
including the degree of agreement with the data. 
The THESEUS simulations for light nuclei are performed for the 
late freeze-out ($\varepsilon_{\rm frz}=$ 0.2 GeV/fm$^3$) for three EoS's. 
Protons are calculated within the conventional 3FD freeze-out with the subsequent UrQMD afterburner.

\begin{figure}[!tbh]
\includegraphics[width=.42\textwidth]{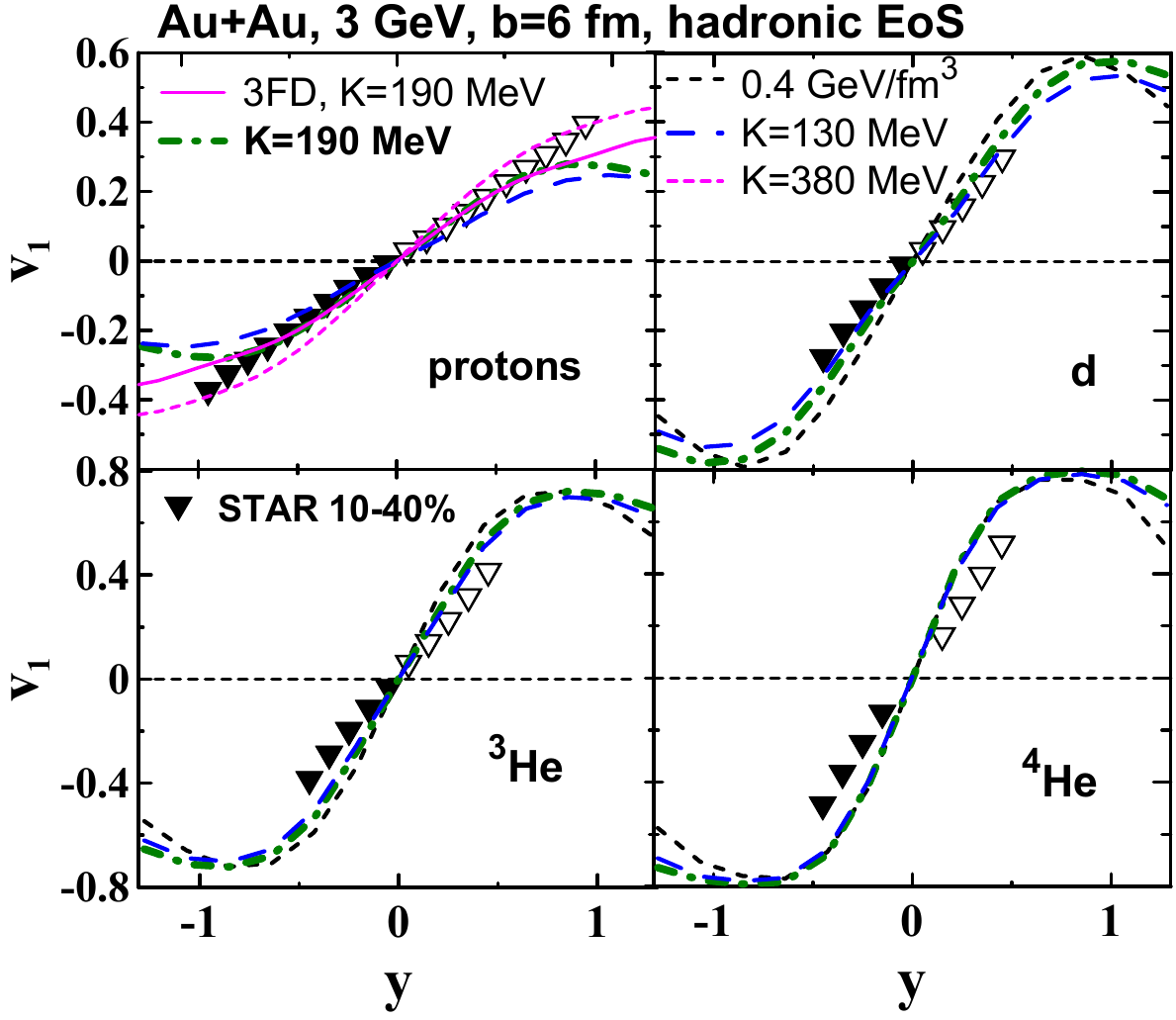}
  \caption{
The same as in Fig. \ref{fig:v1-STAR-fxt-3GeV-fragments_frz02} but for various versions of 
hadronic EoS: the standard hadronic EoS ($K=$ 190 MeV) and very soft hadronic EoS ($K=$ 130 MeV). 
Results of the THESEUS simulations with the 
late freeze-out ($\varepsilon_{\rm frz}=$ 0.2 GeV/fm$^3$) and   
conventional 3FD freeze-out ($\varepsilon_{\rm frz}=$ 0.4 GeV/fm$^3$)  
are displayed for light nuclei. 
Protons are calculated within the conventional 3FD freeze-out with the subsequent UrQMD afterburner.
The proton $v_1$ within the 3FD model, i.e. before the UrQMD afterburner, is also presented 
for the standard EoS ($K=$ 190 MeV, thin solid line) and stiff EoS 
($K=$ 380 MeV, thin short-dashed line).  
}	
    \label{fig:v1-STAR-fxt-gas3GeV-fragments_frz02}
\end{figure}

The directed flow turns out to be independent of the used EoS, which again suggests that the dynamics
is dominated by the hadronic phase. The calculated results almost perfectly 
(except for very forward and backward rapidities) reproduce the experimental 
proton directed flow \cite{STAR:2021yiu}. Agreement with the data \cite{STAR:2021ozh}
is getting worse with increase of atomic number of light nucleus. 
If the calculated midrapidity slope of the directed flow is only slightly steeper than 
the experimental one for deuterons, for $^4$He it is already noticeably steeper.

To check if this disagreement is related to the above observed preference of the 
conventional freeze-out for the $^4$He, see Fig. \ref{fig:dNdy_All_3AGeVmix_tph_gas_frz02_frz04}, 
we present the results with conventional 3FD freeze-out in Fig. \ref{fig:v1-STAR-fxt-gas3GeV-fragments_frz02}. 
As seen, the $^4$He flow is independent of the type of the freeze-out while the flow slopes of lighter nuclei
become only slightly steeper at the conventional freeze-out.  

Stiffness of the hadronic EoS with the 3FD model can be easily changed. The stiffness is characterized by 
incompressibility of nuclear matter that is conventionally defined as 
\begin{eqnarray}
K = 9n_0^2 \frac{d^2}{d n^2} \left( \frac{\varepsilon (n,T=0)}{n} \right)_{n=n_0},  
\end{eqnarray}
where $\varepsilon (n,T=0)$ is the energy density of the nuclear matter at zero temperature ($T=0$) 
as a function of the baryon density ($n$), $n_0$ is the normal nuclear density. 
The conventionally used hadronic EoS is characterized by $K=$ 190 MeV. This is a quite soft EoS. 
It is very similar (but not identical) to the EoS of the hadronic phase 
in the 1PT and crossover EoS's \cite{Toneev06}. To study the effect of the EoS stiffness on the directed flow, 
we present results for a very soft hadronic EoS ($K=$ 130 MeV) 
in Fig. \ref{fig:v1-STAR-fxt-gas3GeV-fragments_frz02}. 
As seen, the $^4$He flow again turns out to be independent of the EoS stiffness. 
The very soft EoS gives slightly better agreement with the data for lighter nuclei
than the conventionally used EoS  
but results in disagreement with the experimental proton flow. 
In Ref. \cite{STAR:2021yiu} it is reported that the stiff hadronic EoS ($K=$ 380 MeV) 
well reproduces the proton directed flow within the UrQMD and JAM models. 
In contrast, our calculation shows that the stiff EoS ($K=$ 380 MeV)
results in too steep slope of the proton flow 
(see the thin short-dashed line in Fig. \ref{fig:v1-STAR-fxt-gas3GeV-fragments_frz02}), 
which leads to even stronger disagreement with flow data for light nuclei. 
Therefore, the conventionally used soft hadronic EoS with $K=$ 190 MeV seems to be the optimal choice. 
This conclusion agrees with that in famous paper \cite{Danielewicz:2002pu} made more than twenty years ago.

The proton $v_1$ within the 3FD model, i.e. before the UrQMD afterburner, is also presented  
in Fig. \ref{fig:v1-STAR-fxt-gas3GeV-fragments_frz02}. The afterburner does not change the 
midrapidity slope of the flow but worsens agreement with the data at very forward and backward
rapidities. The proton flow at the late freeze-out and without afterburner (not displayed in 
Fig. \ref{fig:v1-STAR-fxt-gas3GeV-fragments_frz02}) is very similar to that at the conventional 
3FD freeze-out and the subsequent afterburner, which once again confirms the correctness of the choice of energy density for the late freeze-out. 

In contrast to the proton flow, the afterburner imitation (i.e. the late freeze-out) does change 
the midrapidity slope of deuterons and $^3$He albeit slightly. However, the $^4$He flow 
is not affected by the late freeze-out.

\subsection{Elliptic flow} 
  \label{Elliptic flow}

\begin{figure}[!tbh]
\includegraphics[width=.46\textwidth]{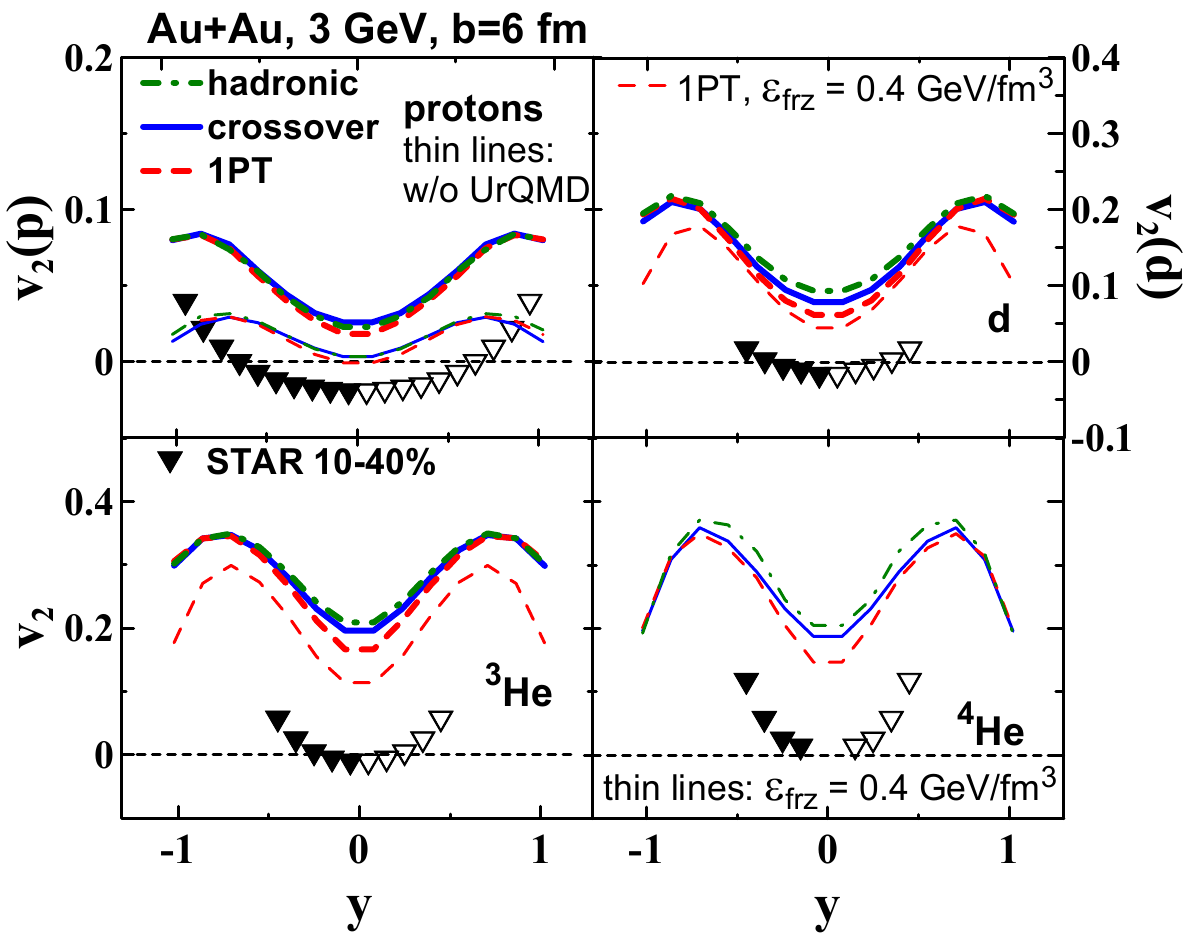}
  \caption{
Elliptic flow of  
protons and light nuclei 
(deuterons, $^3$He and $^4$He) as function of rapidity in semicentral ($b=$ 6 fm)    
Au+Au collisions at collision energy of $\sqrt{s_{NN}}=$ 3 GeV. 
Results are calculated 
with hadronic, 1PT and crossover EoS's. 
Results of the THESEUS simulations with three EoS's are displayed
for deuterons and $^3$He at the late freeze-out ($\varepsilon_{\rm frz}=$ 0.2 GeV/fm$^3$)
and for the $^4$He nuclei at the 
conventional 3FD freeze-out ($\varepsilon_{\rm frz}=$ 0.4 GeV/fm$^3$). 
The conventional 3FD freeze-out with the 1PT EoS is also presented for for deuterons and $^3$He.
Protons are calculated within the conventional 3FD freeze-out with the subsequent UrQMD afterburner.
The proton $v_2$ flows before the UrQMD afterburner are also presented. 
STAR data are from Refs. \cite{STAR:2021ozh,STAR:2021yiu}.
Full symbols display measured experimental points, whereas the open ones are those reflected with respect
to the midrapidity.}	
    \label{fig:v2-STAR-fxt-3GeV-fragments_frz02}
\end{figure}

The calculated elliptic flow of protons and light nuclei (deuterons, $^3$He and $^4$He) 
as function of rapidity in semicentral ($b=$ 6 fm)    
Au+Au collisions at collision energy of $\sqrt{s_{NN}}=$ 3 GeV is presented in 
Fig. \ref{fig:v2-STAR-fxt-3GeV-fragments_frz02}. 
The results are compared with STAR data \cite{STAR:2021ozh,STAR:2021yiu}.
Results for tritons are again omitted because they are very similar to those for $^3$He. 
The THESEUS simulations for deuterons and $^3$He are performed for 
the late freeze-out ($\varepsilon_{\rm frz}=$ 0.2 GeV/fm$^3$), while for $^4$He,  
the conventional 3FD freeze-out ($\varepsilon_{\rm frz}=$ 0.4 GeV/fm$^3$) in view of its 
preference for $p_T$ spectra and $y$ distributions, see sect.  \ref{Bulk Observables}. 
Protons are calculated within the conventional 3FD freeze-out with the subsequent UrQMD afterburner.
For comparison, the light-nuclei flow with 
the conventional 3FD freeze-out for the 1PT EoS
and the proton $v_2$ flow before the UrQMD afterburner are also demonstrated.

\begin{figure}[!tbh]
\includegraphics[width=.46\textwidth]{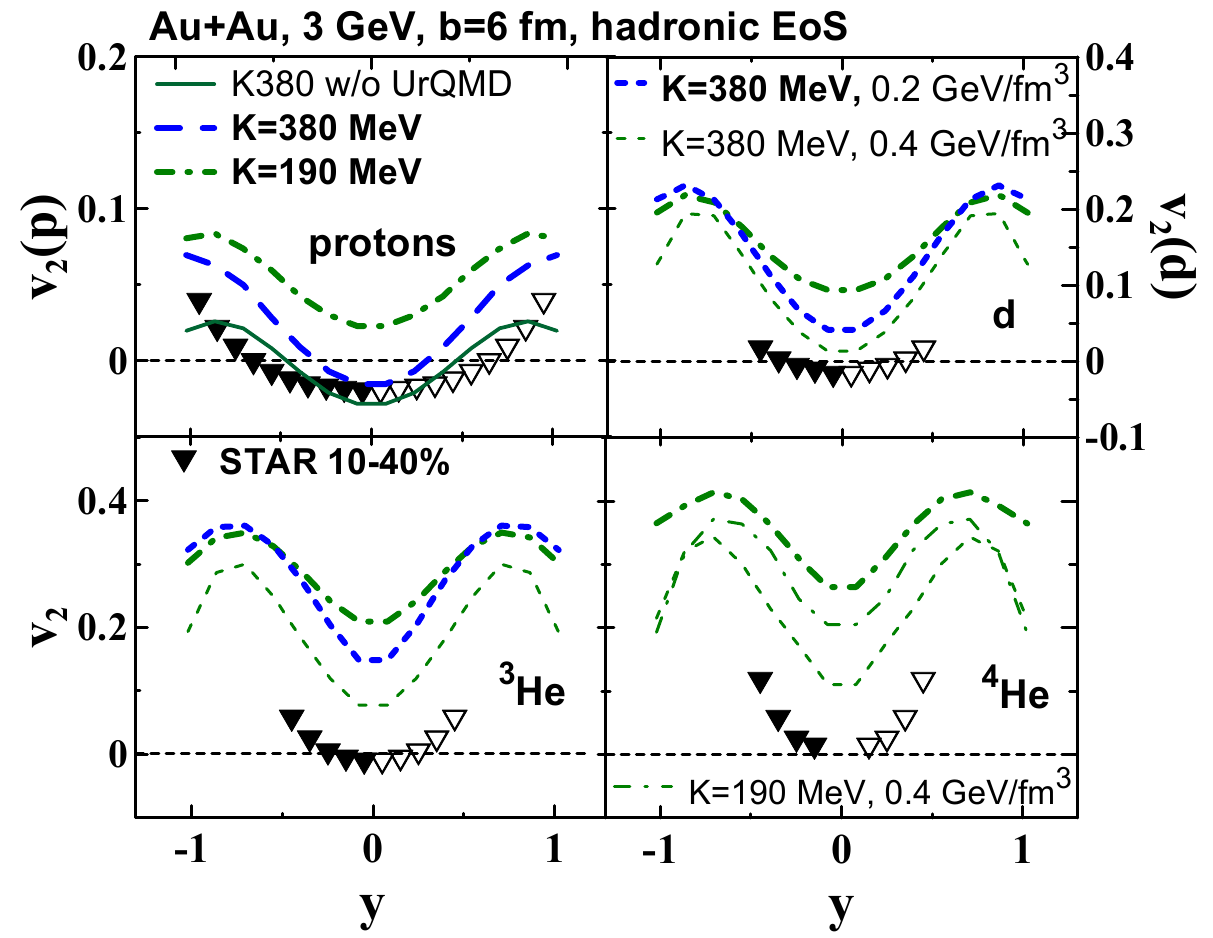}
  \caption{
The same as in Fig. \ref{fig:v2-STAR-fxt-3GeV-fragments_frz02} but for various versions of the hadronic EoS. 
Results with the standard hadronic EoS ($K=$ 190 MeV) are displayed 
by bold dash-dotted lines for the late 
freeze-out ($\varepsilon_{\rm frz}=$ 0.2 GeV/fm$^3$) for all light nuclei 
and by thin dash-dotted line for the standard 3FD freeze-out ($\varepsilon_{\rm frz}=$ 0.4 GeV/fm$^3$) 
for $^4$He nuclei. 
Results of calculations with the stiff hadronic EoS ($K=$ 380 MeV) are also presented: 
for deuterons and $^3$He at the late freeze-out and for all nuclei at the standard 3FD freeze-out. 
Protons are calculated within the conventional 3FD freeze-out with the subsequent UrQMD afterburner.
The proton $v_2$ before the UrQMD afterburner is also presented for the stiff EoS ($K=$ 380 MeV)
by  solid line.  
}	
    \label{fig:v2-STAR-fxt-gas3GeV-fragments_frz02}
\end{figure}

As seen from Fig. \ref{fig:v2-STAR-fxt-3GeV-fragments_frz02}, the calculated elliptic flow considerably 
overestimates the data \cite{STAR:2021ozh,STAR:2021yiu}. Even the $v_2$ sign is different in the midrapidity 
region. The afterburner (for protons) and the late freeze-out (for light nuclei) even worsen agreement 
with the data. The large disagreement of light-nuclei elliptic flow with the data \cite{STAR:2021ozh}
does not mean that the thermodynamic approach fails to describe this flow. This only means that the 3FD model 
has troubles describing the elliptic flow of protons, 
which transform into even biger troubles for light nuclei.

As stated in Refs. \cite{STAR:2021ozh,STAR:2021yiu,Danielewicz:2002pu}, a stiff EoS is needed 
for describing the elliptic flow. Therefore, we performed simulations with stiff hadronic EoS ($K=$ 380 MeV),
see Fig. \ref{fig:v2-STAR-fxt-gas3GeV-fragments_frz02}. The results became closer to the data. 
The proton elliptic flow is even reproduced in the midrapidity. However, the overall disagreement with 
data remains. The afterburner (late freeze-out) still worsens the agreement with data.

Note that dips and even negative values of $v_2$ in the midrapidity are consequences of  
the squeeze-out effect \cite{Sorge:1996pc,Danielewicz:1998vz,Ivanov:2014zqa}, 
which result from blocking of the expanding central blob by the spectator matter.
The squeeze-out is a characteristic feature of moderately relativistic collisions, in which 
the expanding central fireball is shadowed by spectators. This shadowing only partially is 
taken into account within the 3FD evolution because the frozen-out matter of the central fireball 
remains to be shadowed even after the freeze-out while in the 3FD model it 
escapes without interacting with spectators. 
The afterburning stage should, in principle, correct this deficiency. 
However, it does not, as we see from 
Figs. \ref{fig:v2-STAR-fxt-gas3GeV-fragments_frz02} and \ref{fig:v2-STAR-fxt-3GeV-fragments_frz02}. 
The reason is that the THESEUS assigns {\em the same time instant} to all produced particles
during the particlization procedure, while different parts of the system are frozen-out at 
{\em different time instants} in 3FD. 
Participants are frozen out earlier than spectators. If the particlization is isochronous,   
the evolution of the frozen-out participants stops untill the spectators also become frozen-out. 
Therefore, we skip the stage of shading the afterburner expansion of the central fireball by spectators still being in the hydrodynamic phase.
The afterburner evolution is switched on only when the spectators also become frozen out. 
The spectators are frozen out when they have already passed the expanding central fireball.
Thus, the shadowing by spectators turns out to be strongly reduced 
after such isochronous particlization compared to what it would be if the entire collision process were 
kinetically treated, as in UrQMD or JAM.  
Apparently, this is the prime reason of failure of the elliptic-flow description. 
A time-extended transition from hydrodynamic evolution to afterburner dynamics would need to take into account the interaction of the kinetic afterburner phase with still hydrodynamically evolved matter. This is a difficult task both technically and conceptually.

\section{Feed-Down from Unstable $^4He$}

\begin{figure}[!tbh]
  \includegraphics[width=.3\textwidth]{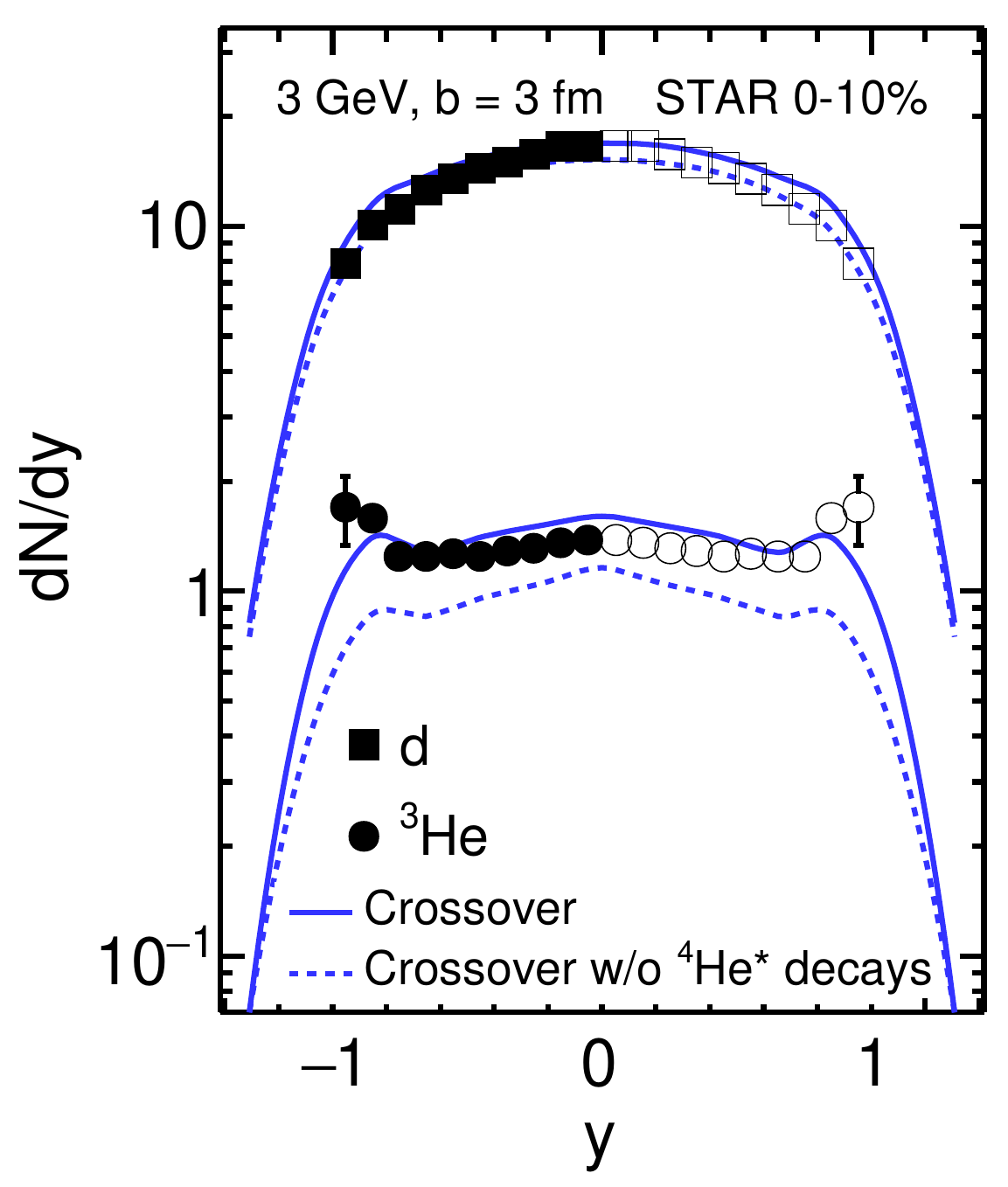}
  \caption{
Rapidity distributions of light nuclei 
(deuterons and $^3$He)  in  central 
Au+Au collisions at collision energy of $\sqrt{s_{NN}}=$ 3 GeV.
Results are calculated in the crossover scenario with  
and without (w/o $^4$He$^*$ decays) the feed-down contributions.  
STAR data are from Ref. \cite{STAR:2023uxk}.
Full symbols display measured experimental points, whereas the open ones are those reflected with respect
to the midrapidity.}	
    \label{fig:dNdy_3AGeVmix_3fm_frz02_d_He3_noHe4dec}
\end{figure}

As found in Ref. \cite{Kozhevnikova:2022wms}, the feed-down contributions from unstable $^4$He
to deuterons are negligibly small, while to tritons and $^3$He are less than 20\% at $\sqrt{s_{NN}}>$ 6 GeV 
in the midrapidity. At the forward/backward rapidities, these contributions are essential even 
at $\sqrt{s_{NN}}>$ 6 GeV. 
It was predicted \cite{Vovchenko:2020dmv}, that feed-down contributions
reach values of the order of 60\% for tritons and $^3$He even in midrapidity at $\sqrt{s_{NN}}=$ 3 GeV. 

Results of our calculations are presented in Fig. \ref{fig:dNdy_3AGeVmix_3fm_frz02_d_He3_noHe4dec}
at the example of the crossover EoS. 
In agreement with Ref. \cite{Vovchenko:2020dmv}, the feed-down contribution amounts $\sim$20\% 
for deuterons and 50--100\% (depending on the rapidity) for $^3$He. 
While the feed-down contribution into the deuteron yield is inessential for the data reproduction, 
it plays important role for $^3$He. Without this feed-down the $^3$He yield is noticably 
underestimated.

The $v_1$ flow of deuterons, tritons and $^3$He turns out to be insensitive to 
the feed-down contributions from unstable $^4$He. 
Without these contributions, the corresponding $v_2$ flows are reduced by $\sim$20\%, 
which, however, does not  essentially change the degree of their agreement with the data. 
The effect on the proton yield and flow is negligible.

\section{Summary}
\label{Summary}

Simulations of the proton and light-nuclei production 
in Au+Au collisions at $\sqrt{s_{NN}}=$ 3 GeV were performed
within the updated THESEUS event generator \cite{Kozhevnikova:2020bdb}. 
The results were compared with recent STAR data \cite{STAR:2023uxk,STAR:2021ozh}.
The updated THESEUS treats the light-nuclei production 
within the thermodynamical approach on the equal basis with hadrons.
The protons (as well as other hadron) are calculated with the standard 3FD freeze-out, 
characterized by the energy density of $\varepsilon_{\rm frz}=$ 0.4 GeV/fm$^3$, 
followed by the UrQMD afterburner. 
The only additional parameter related to the light nuclei is the energy density of 
the late freeze-out that imitates the afterburner stage because the light nuclei 
do not participate in the UrQMD evolution.

In fact, the freeze-out parameter is required in both the thermodynamical and coalescence approaches. 
Both approaches are inapplicable at very early freeze-out, when 
inter-particle spacing in a fireball is less than the inter-nucleon distance in a light nucleus.  
At later freeze-out, the light-nuclei yields crucially depend on the freeze-out conditions in both approaches. 
The coalescence allows fine tuning of the light-nuclei production after the freeze-out
by means of coalescence parameters. In contrast, within the thermodynamical approach, 
the light-nuclei observables are solely determined by the applied freeze-out in the absence of the afterburner. 
Dynamical treatment of the light nuclei at the afterburner stage 
\cite{Oliinychenko:2018ugs,Staudenmaier:2021lrg,Sun:2021dlz,Sun:2022xjr}
would essentially weaken this strong dependence on the freeze-out.

It was found that the late freeze-out characterized by 
 $\varepsilon_{\rm late \: frz}=$ 0.2 GeV/fm$^3$ is preferable for deuterons, tritons, and $^3$He.  
This is precisely the same value of $\varepsilon_{\rm late \: frz}$ that was found in 
Ref. \cite{Kozhevnikova:2022wms} at higher collision energies. 
Remarkably, the $^4$He yield and $p_T$ spectra are better reproduced with the standard 3FD freeze-out. 
This suggests that the $^4$He nuclei better survive 
in the afterburner stage because they are more spatially compact and tightly bound objects. 
This is an argument in favor of dynamical treatment of light nuclei.

Results of simulations with different EoS's (with and without transition to the quark-gluon phase)
indicated that the dynamics is determined by the hadronic phase. 
The calculated results revealed not perfect, but a good reproduction of 
the data on bulk observables of the light nuclei.
The calculated proton directed flow almost perfectly 
(except for very forward and backward rapidities) reproduces the experimental 
flow \cite{STAR:2021yiu}. Agreement with the data on the directed flow \cite{STAR:2021ozh}
becomes worse with increase of atomic number of light nucleus. 
If the calculated midrapidity slope of the directed flow is only slightly steeper than 
the experimental one for deuterons, for $^4$He it is already noticeably steeper.

The model failed to properly describe the data on the elliptic flow of both protons and light nuclei. 
We attribute this to shortcomings of the transition from the 3FD evolution to the UrQMD afterburner, 
which prevents us from proper description of the squeeze-out effect. 
The squeeze-out results from shadowing of the expanding central fireball by spectators.
This shadowing only partially 
taken into account within the 3FD evolution because the frozen-out matter of the central fireball 
remains to be shadowed even after the freeze-out while in the 3FD model it 
escapes without interacting with spectators. 
The afterburning stage should, in principle, correct this deficiency but it does not. 
The reason is that the THESEUS assigns  the same time instant to all produced particles
during the particlization procedure, while different parts of the system get frozen-out at 
different time instants in 3FD. 
The shadowing by spectators is strongly reduced 
after such isochronous particlization because the participants and spectators turn out to be well separated in 
thus constructed pre-afterburner configuration.

We also studied the feed-down contributions from unstable $^4$He and possible in-medium effects. 
As found, the feed-down contribution amounts $\sim$20\% 
for deuterons and 50--100\% (depending on the rapidity) for tritium and $^3$He. 
While the feed-down contribution into the deuteron yield is inessential for the data reproduction, 
it plays important role for tritium and $^3$He. Without this feed-down the tritium and $^3$He 
yields are noticably underestimated. 
The medium effect accordingly to Ref. \cite{Ropke:2014fia} turned out to be too strong and  
resulted in strong disagreement with data. Apparently this is because the 
relativistic mean-field model \cite{Ropke:2012qv}, underlying these medium corrections, 
has been parametrized to reproduce low-energy nuclear phenomena. This parametrization is simply inapplicable 
to highly excited nuclear matter.

Development of new hybrid model called MUFFIN (MUlti Fluid simulation for Fast IoN collisions) 
was recently announced in Ref. \cite{Cimerman:2023hjw}. This is a next-generation hybrid three-fluid
model for simulating heavy-ion collisions at energies from few to few tens of GeV.
Several methodical and conceptual improvements, as compared with THESEUS, are introduced in MUFFIN. 
An important conceptual improvement is the inclusion of initial state fluctuations, which are important when considering the collective flow and allow study of fluctuations associated with the CEP. 
The afterburner based on SMASH will make it possible to describe the afterburner evolution of at least deuterons. 
This could resolve some of the aforementioned problems in THESEUS.

\begin{acknowledgments}

We are sincerely grateful to Iurii Karpenko and David Blaschke 
who made enormous contributions early on in the implementation of this project.  
Fruitful discussions with  D.N. Voskresensky are gratefully acknowledged.
This work was carried out using computing resources of the federal collective usage center ``Complex for simulation and data processing for mega-science facilities'' at NRC "Kurchatov Institute" \cite{ckp.nrcki.ru}.
and computing resources of the supercomputer "Govorun" at JINR \cite{govorun}. 

\end{acknowledgments}


\end{document}